\documentclass[structabstract]{aa}
\usepackage{amssymb}
\usepackage{natbib}
\usepackage{txfonts}
\usepackage{graphicx}
\usepackage{dcolumn} 
\newcommand{\be}{\begin{equation}}
\newcommand{\ee}{\end{equation}}
\newcommand{\ba}{\begin{eqnarray}}
\newcommand{\ea}{\end{eqnarray}}
\newcommand{\beqa}{\begin{eqnarray}}
\newcommand{\eeqa}{\end{eqnarray}}

\begin{document}

\title{Weak gravitational lensing effects on cosmological parameters and dark energy from gamma-ray bursts}

\author{F. Y. Wang$^{1,2}$ and Z. G. Dai$^{1,2}$}
\institute{Department of Astronomy, Nanjing University, Nanjing
210093, China \and Key laboratory of Modern Astronomy and
Astrophysics (Nanjing University), Ministry of Education, Nanjing
210093, China}

   \offprints{F. Y. Wang (fayinwang@nju.edu.cn) and Z. G. Dai (dzg@nju.edu.cn)}

   \authorrunning{Wang \& Dai}
\titlerunning{Gravitational Lensing Effects on GRBs}

\abstract {Gamma-ray bursts (GRBs) are attractive objects for
constraining the nature of dark energy in a way complementary to
other cosmological probes, especially at high redshifts. However,
the apparent magnitude of distant GRBs can be distorted by the
gravitational lensing from the density fluctuations along the line
of sight.} {We investigate the gravitational lensing effect on the
cosmological parameters and dark energy equation of state from
GRBs.} {We first calibrated the GRB luminosity relations without
assuming any cosmological models. The luminosity distances of
low-redshift GRBs were calibrated with the cosmography method using
a latest type Ia supernova (SNe Ia) sample. The luminosity distances
of high-redshift GRBs were derived by assuming that the luminosity
relations do not evolve with redshift. Then we investigated the
non-Gaussian nature of the magnification probability distribution
functions and the magnification bias of the gravitational lensing.}
{We find that the gravitational lensing has non-negligible effects
on the determination of cosmological parameters and dark energy. The
gravitational lensing shifts the best-fit constraints on
cosmological parameters and dark energy. Because high-redshift GRBs
are more likely to be reduced, the most probable value of the
observed matter density $\Omega_M$ is slightly lower than its actual
value. In the $\Lambda$CDM model, we find that the matter density
parameter $\Omega_M$ will shift from $0.30$ to $0.26$ after
including the gravitational lensing effect. The gravitational
lensing also affects the dark energy equation of state by shifting
it to a more negative value. We constrain the dark energy equation
of state out to redshift $z\sim 8$ using GRBs for the first time,
and find that the equation of state deviates from $\Lambda$CDM at
the $1\sigma$ confidence level, but agrees with $w=-1$ at the
$2\sigma$ confidence level.} {}


   \maketitle

\section{Introduction}\label{sec:introduction}
Our Universe is currently in accelerating expansion. This is
revealed by observations of high-redshift type Ia supernovae (SNe
Ia) (Riess et al. 1998; Perlmutter et al. 1999), cosmic microwave
background (CMB) fluctuations (Spergel et al. 2003, 2007), and the
large-scale structure (LSS) (Tegmark et al. 2006). These
observations suggest that the composition of the universe may
consist of an extra component such as dark energy, or that the
equations governing gravity may need a variation to explain the
acceleration of the universe at the present epoch.

In order to measure the expansion history of our Universe, we need
the Hubble diagram of standard candles. Type Ia supernovae that have
played an important role in constraining cosmological parameters are
the well-known standard candles. Unfortunately, it is difficult to
observe SNe Ia at $z>1.7$, even with excellent space-based projects
such as SNAP (Aldering et al. 2004). They cannot provide any
information on the cosmic expansion beyond redshift 1.7. On the
other hand, gamma-ray bursts (GRBs) are ideal candidates for this
investigation. The high luminosities of GRBs make them detectable
out to the edge of the visible universe (Bromm \& Loeb 2002, 2006).
The farthest GRB is GRB 090429B at $z=9.4$ (Cucchiara et al. 2011).
Schaefer (2007) compiled 69 GRBs to simultaneously use five
luminosity indicators, which are relations of $\tau_{\rm lag}-L$
(Norris, Marani \& Bonnell 2000), $V-L$ (Fenimore \& Ramirez-Ruiz
2000), $E_{\rm peak}-L$ (Schaefer et al. 2003), $E_{\rm
peak}-E_{\gamma}$(Ghirlanda, Ghisellini \& Lazzati 2004), and
$\tau_{\rm RT}-L$ (Schaefer 2007). Here the time lag ($\tau_{\rm
lag}$) is the time shift between the hard and soft light curves, $L$
is the peak luminosity of a GRB, the variability $V$ of a burst
denotes whether its light curve is spiky or smooth and it can be
obtained by calculating the normalized variance of an observed light
curve around a smoothed version of that light curve (Fenimore \&
Ramirez- Ruiz 2000), $E_{\rm peak}$ is the photon energy at which
the $\nu F_{\nu}$ spectrum peaks, $E_{\gamma}=(1-\cos\theta_j)E_{\rm
\gamma, iso}$ is the collimation-corrected energy of a GRB, and the
minimum rise time ($\tau_{\rm RT}$) in the gamma-ray light curve is
the shortest time over which the light curve rises by half of the
peak flux of the pulse. Qi \& Lu (2010) found a new correlation in
the X-ray band of GRB afterglows. After calibrating with luminosity
relations, GRBs may be used as standard candles to provide
information on the cosmic expansion at high redshift and, at the
same time, to tighten the constraints on cosmic expansion at low
redshift (Dai et al. 2004; Ghirlanda et al. 2004; Friedman \& Bloom
2005; Liang \& Zhang 2005, 2006; Wang \& Dai 2006; Schaefer 2007a;
Wright 2007; Wang, Dai \& Zhu 2007; Wang 2008; Qi, Wang \& Lu
2008a,b; Liang et al. 2008; Amati et al. 2008; Cardone et al. 2009;
Liang et al. 2009; Qi, Lu \& Wang 2009; Izzo et al. 2009). Gamma-ray
bursts can also potentially probe the cosmographic parameters to
distinguish between dark energy and modified gravity models (Wang,
Dai  \& Qi 2009a, b; Vitagliano et al. 2010; Capozziello \& Izzo
2008; Xia, et al. 2011). Samushia \& Ratra (2010) also derived
constraints on the $\Lambda$CDM and XCDM models using a smaller set
of 69 GRBs analyzed in two different ways, following Schaefer (2007)
and Wang (2008). They found that GRB data disfavor standard
$\Lambda$CDM at a 1$\sigma$ confidence level, and is consistent with
this model at a 2$\sigma$ confidence level. They also found that the
two techniques give somewhat different cosmological constraints.
This means that model-independent calibration method and more GRB
data are needed to improve this situation. We will use the
cosmographic parameters to calibrate the GRB luminosity
correlations. More recently, Wang, Qi \& Dai (2011) enlarged the GRB
sample and put constraints on cosmological parameters and equation
of state of dark energy. In this paper, we use this GRB sample which
includes 116 GRBs.

However, the gravitational lensing by random fluctuations in the
intervening matter distribution induces a dispersion in GRB
brightness (Oguri \& Takahashi 2006; Schaefer 2007), degrading their
value as standard candles as well as SNe Ia (Holz 1998). Gamma-ray
bursts can be magnified (or reduced) by the gravitational lensing
produced by the structure of the Universe. The gravitational lensing
has sometimes a great impact on high-redshift GRBs. First, the
probability distribution functions (PDFs) of gravitational lensing
magnification have much higher dispersions and are markedly farther
from the Gaussian distribution more remarkably (Valageas 2000a; Wang
et al. 2002; Ogri \& Takahashi 2006). Second, there is effectively a
threshold for the detection in the burst apparent brightness. With
gravitational lensing, bursts just below this threshold might be
magnified in brightness and detected, whereas bursts just beyond
this threshold might be reduced in brightness and excluded. Schaefer
(2007) considered the gravitational lensing biases and the Malmquist
biases of 69 GRBs. He found that the gravitational lensing and
Malmquist biases are much smaller than the intrinsic error bars. Our
method differs in two ways from the one used in Schaefer (2007).
First, we use the latest GRB sample that includes 116 GRBs. Second,
we calculate the distance dispersions from the universal probability
distribution function (UPDF) of the gravitational lensing
amplification (Wang et al. 2002; Wang 2005). But Schaefer (2007)
used the method of Gonzalez \& Faber (1997), which relies on the
poorly known luminosity function and number densities of GRBs.

We explore the gravitational lensing effects on constraints of
cosmological parameters and equation of state from GRBs. We focus on
the non-Gaussian nature of magnification probability distribution
functions and the magnification bias of the gravitational lensing.
The structure of this paper is arranged as follows: in section 2 we
calibrate the luminosity relations of GRBs in a cosmological
model-independent way. The constraints on the cosmological
parameters and dark energy including gravitational lensing are
presented in section 3. In section 4 we present model-independent
constraints on the dark energy equation of state to $z\sim8$
including the weak gravitational effect. In section 5 we summarize
our findings and give a brief discussion.

\section{Calibration of the luminosity relations of GRBs}\label{sec:relation}
In this section we calibrate the GRB luminosity relations with
cosmographic parameters. The cosmographic parameters are
cosmology-independent. The only assumption is the basic symmetry
principles (the cosmological principle) that the universe can be
described by the Friedmann-Robertson-Walker metric.

The expansion rate of the Universe can be written in terms of the
Hubble parameter, $H=\dot{a}/a$, where $a$ is the scale factor and
$\dot{a}$ is its first derivative with respect to time. Because we
know that $q$ is the deceleration parameter, related to the second
derivative of the scale factor, $j$ is the so-called ``jerk'' or
statefinder parameter, related to the third derivative of the scale
factor, and $s$ is the so-called ``snap'' parameter, which is
related to the fourth derivative of the scale factor. These
quantities are defined as
\begin{equation}
q=-\frac{1}{H^2}\frac{\ddot{a}}{a};
\end{equation}
\begin{equation}
j=\frac{1}{H^3}\frac{\dot{\ddot{a}}}{a};
\end{equation}
\begin{equation}
s=\frac{1}{H^4}\frac{\ddot{\ddot{a}}}{a}.
\end{equation}
The deceleration, jerk and snap parameters are dimensionless, and a
Taylor expansion of the scale factor around $t_0$ provides
\begin{eqnarray}
a(t)=a_0 \left\{1+H_0(t-t_0)-\frac{1}{2}q_0H_{0}^{2}(t-t_0)^2
+\frac{1}{3!}j_0H_{0}^{3}(t-t_0)^3\right.
\nonumber\\
+\frac{1}{4!}s_0H_{0}^{4}(t-t_0)^4+O[(t-t_0)^5]\},
\end{eqnarray}
and so the luminosity distance \beqa
d_L={{c}\over{H_0}}\left\{z+{{1\over2}(1-q_0)}z^2-{{1}\over{6}}
\left(1-q_0-3q_0^2+j_0\right)z^3\right.
\nonumber\\
+{{1}\over{24}}\left[2-2q_0-15q_{0}^{2}-15q_0^3+5j_0 +10q_0
j_0+s_0\right]z^4 +O(z^5)\}, \eeqa (Visser 2004). Catto\"{e}n \&
Visser (2007) pointed out that it is useful to recast $d_L$ as a
function of an improved parameter $y=z/(1+z)$ and constrained the
cosmographic parameters using SNe Ia data. In this way, because
$z\in(0,\infty)$ mapped into $y\in(0,1)$, the luminosity distance
with the fourth order in the $y$-parameter becomes
\beqa
\begin{array}{l}
 d_L (y) = \frac{c}{{H_0 }}\{y - \frac{1}{2}(q_0  - 3)y^2  + \frac{1}{6}\left[ {12 - 5q_0  - (j_0  + \Omega _0 )} \right]y^3  \\
  + \frac{1}{{24}}\left[ {60 - 7j_0  - 10\Omega _0  - 32q_0  + 10q_0 j_0  + 6q_0 \Omega _0  + 21q_0^2  - 15q_0^3  + s_0 } \right]y^4  \\
  + O(y^5 )\}, \\
 \end{array}\eeqa where $\Omega_0=1+kc^2/H_0^2a^2(t_0)$ is the total energy
density.

\begin{figure}
\includegraphics[width=9cm]{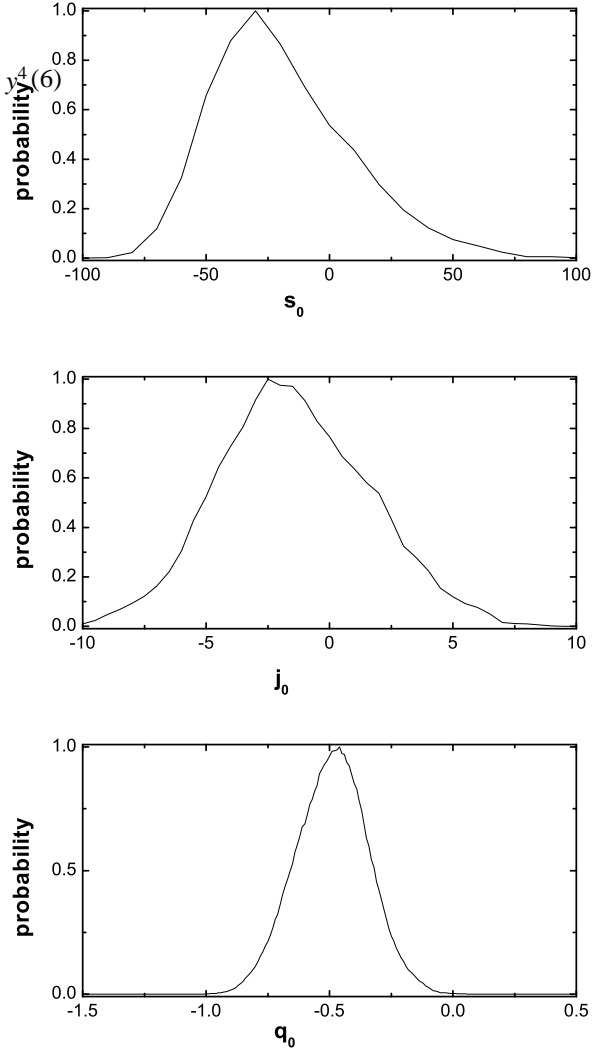}
\caption{The probability distribution of $q_0$, $j_0$ and $s_0$.
 \label{cosmo}}
\end{figure}

We used the Union2 (Amanullah et al. 2010) dataset to fit the
cosmographic parameters. We set the Hubble parameter
$H_0=70$~km/s/Mpc. The likelihood function for $q_0, j_0, s_0$ can
be determined from $\chi^2$ statistics
\begin{equation}
\chi^2=A-\frac{B^2}{C}+\ln\left(\frac{C}{2\pi}\right)~,
\end{equation}
where
\begin{eqnarray}
A=\sum_i\frac{(\mu^{\rm data}-\mu^{\rm th})^2}{\sigma^2_i}~,~~~
B=\sum_i\frac{\mu^{\rm data}-\mu^{\rm th}}{\sigma^2_i}~,~~~
C=\sum_i\frac{1}{\sigma^2_i}~.
\end{eqnarray}
In the calculation, we used Markov chain Monte Carlo techniques. A
Markov chain with samples on the order of $10^6$ is generated
according to the likelihood function and then properly burned in and
thinned to derive statistics of the parameters of interest $q_0$,
$j_0$ and $s_0$. The best-fitting results are
\begin{equation}
q_0=-0.46\pm0.15~~,~~~j_0=-1.85\pm2.25,~~~s_0=-25.53\pm35.30\nonumber.
\label{bestfit}
\end{equation}
The probability distribution of $q_0$, $j_0$ and $s_0$ are shown in
Fig.~\ref{cosmo}. Our results are consistent with those in
Vitagliano et al. (2009), but our results are tighter. Gao et al.
(2010) and Capozziello \& Izzo (2010) used the Union dataset to
calibrate the correlations of GRBs. Izzo, Luongo \& Capozziello
(2010) also used the Union2 dataset to constrain the cosmography
parameters.

\begin{table}[h!!!]
\small \caption[]{Calibration results of five GRBs luminosity
relations.}\label{Table 1} \tabcolsep 4mm
 \begin{tabular}{cccc}
  \hline\noalign{\smallskip}
     relations & a & b & $\sigma_{\rm int} $  \\ \hline
      $L-\tau_{\rm lag}$ \phantom{ss}  & $52.01 \pm 0.19$ &$-0.83 \pm 0.02 $&0.52    \\
       $L-V $\phantom{sssss}   &\phantom{ss}  $51.29\pm0.32   \phantom{ss}  $    & $0.85\pm0.32 \phantom{ss} $    &0.84\\
    $L-E_{\rm peak}$\phantom{ss}   & $51.79 \pm 0.16$ &$1.41 \pm 0.06 $&0.62   \\
     $L-\tau_{\rm RT}$ \phantom{sss}  & $52.72 \pm 0.23$ &$-1.48 \pm 0.26 $&0.58   \\
    $E_{\gamma}-E_{\rm peak}$  & $50.63 \pm 0.10$ &$1.62 \pm 0.04 $&0.18   \\
  \noalign{\smallskip}\hline
\end{tabular}
\end{table}

\begin{figure}
\includegraphics[width=9cm]{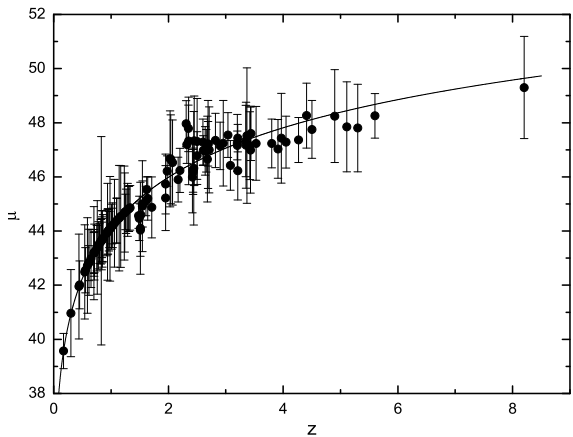}
\caption{Calibrated Hubble diagram of GRBs. The line shows the
distance-redshift relation in a flat $\Lambda$CDM ($\Omega_M=0.3$).
 \label{hubble}}
\end{figure}

We used the latest GRB sample and five luminosity relations in Wang
et al. (2011). We first used the best-fitting cosmography parameters
(see Eq.\ref{bestfit}) to calculate the luminosity distance of GRBs
at $z\leq 1.4$. Then we fitted the five luminosity relations at
$z\leq 1.4$. The calibrated relations are shown in Table\ref{Table
1}. The intrinsic scatter of the $L-V$ relation is too large,
therefor we discard it below. The calibrated luminosity relations
are completely cosmology-independent. Wang et al. (2011) established
that these relations do not evolve with redshift and are valid at
$z>1.40$. The luminosity or energy of a GRB can be calculated at
high redshifts. In this way the luminosity distances and distance
modulus can be obtained. After obtaining the distance modulus of
each burst with one of these relations, we used the same method as
Schaefer (2007) to calculate the real distance modulus,
\begin{equation}
\mu_{\rm fit}=(\sum_i \mu_i/\sigma_{\mu_i}^2)/(\sum_i
\sigma_{\mu_i}^{-2}),
\end{equation}
where the summation runs from $1-4$ over the relations with
available data, $\mu_i$ is the best-estimated distance modulus from
the $i$-th relation, and $\sigma_{\mu_i}$ is the corresponding
uncertainty. The uncertainty of the distance modulus for each burst
is
\begin{equation}
\sigma_{\mu_{\rm fit}}=(\sum_i \sigma_{\mu_i}^{-2})^{-1/2}.
\end{equation}
The calibrated GRB Hubble diagram is shown in Fig.~\ref{hubble}.

\section{Magnification probability distribution of gravitational lensing}

\begin{figure}
\includegraphics[width=9cm]{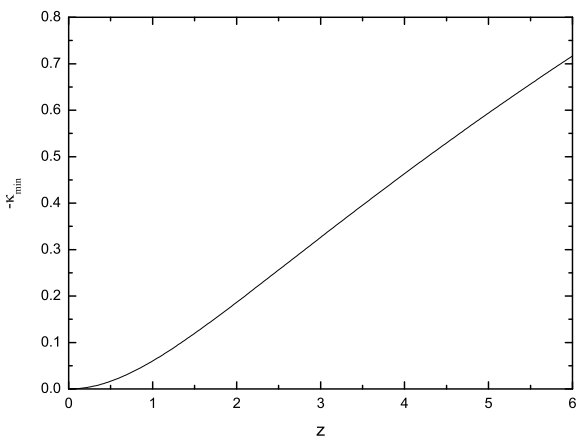}
\caption{The values of $-\kappa_{min}$ in the flat $\Lambda$CDM
($\Omega_M=0.3$).
 \label{kmin}}
\end{figure}

\begin{figure}[htbp]
\includegraphics[width=9cm]{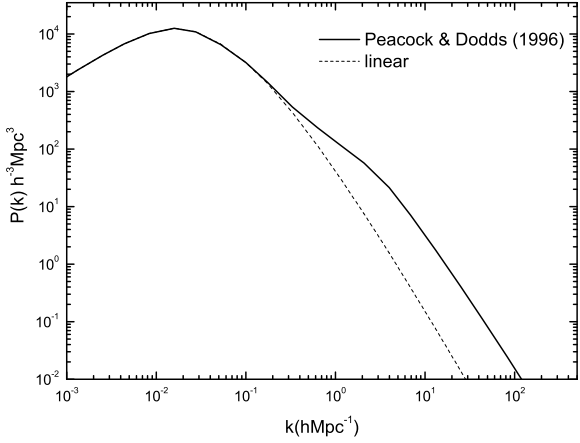}
\caption{Linear and the nonlinear matter power spectrum at $z=1$. We
used the matter transfer function from Bardeen et al. (1986). The
parameters are $\Omega_M=0.27$, $h=0.7$, $\Omega_b=0.04$ and
$\sigma_8=0.96$ (from the WMAP seven-year results).
 \label{mattersepectrum}}
\end{figure}

Owing to the gravitational lensing, the convergence $\kappa$ is
given by (Bernardeau et al. 1997; Kaiser 1998) \be
\kappa=\frac{3}{2}\, \Omega_M \int_0^{\chi_s} \mathrm{d}\chi\,
w(\chi, \chi_s)\, \delta(\chi). \ee Here $\chi$ is the comoving
distance ($\chi_s$ corresponds to the redshift $z_s$ of the source),
\ba \mathrm{d}\chi &=& \frac{cH_0^{-1}\, dz}{
\sqrt{\Omega_{\Lambda}+ \Omega_k
 (1+z)^2+ \Omega_M (1+z)^3}},\nonumber\\
w(\chi,\chi_s) &=& \frac{H_0^2}{c^2}\, \frac{ {\cal{D}}(\chi)\,
{\cal{D}}(\chi_s-\chi)}
{{\cal{D}}(\chi_s)} \, (1+z),\nonumber\\
{\cal{D}}(\chi) &=& \frac{cH_0^{-1}}{\sqrt{|\Omega_k|}}\, {\rm
sinn}\left(\sqrt{|\Omega_k|} \, \frac{H_0}{c}\chi\right),\nonumber
\ea and where $\Omega_k=1-\Omega_M-\Omega_\Lambda$. The density
contrast $\delta \equiv (\rho-\bar{\rho})/\bar{\rho}$. We can see
from Eq.(10) that there exists a minimum value of the convergence:
\be \label{eq:kappamin} \kappa_{min}= -\frac{3}{2}\, \Omega_M
\int_0^{\chi_s} \mathrm{d}\chi\, w(\chi, \chi_s). \ee Accordingly
the minimum magnification factor is
$\tau_{min}=1/(1-\kappa_{min})^2$. The plot of $\kappa_{min}$ is
shown in Fig.~\ref{kmin}. We can derive the magnification PDF for an
arbitrary cosmological model by using the universal probability
distribution function (UPDF) of the gravitational lensing
amplification (Wang et al. 2002; Wang 2005). The UPDF can be fitted
to the stretched Gaussian (Wang et al. 2002), \be \label{eq:P(eta)}
g(\eta|\xi_{\eta})=C_{norm}\, \exp\left[ -\left( \frac{\eta-
\eta_{peak}} {w \,\eta^q} \right)^2 \right], \ee
where $C_{norm}$, ${\eta}_{peak}$, $w$, and $q$ depend on
$\xi_{\eta}$ and are independent of $\eta$. The parameter $\eta$ is
defined by (Valageas 2000a, 2000b) \be \label{eq:eta} \eta \equiv
\frac{ \tau-\tau_{min}}{1-\tau_{min}}=
1+\frac{\kappa}{|\kappa_{min}|} =\frac{\int_0^{\chi_s}
\mathrm{d}\chi\, w(\chi, \chi_s)\, \left(\rho/\bar{\rho}\right)}
{\int_0^{\chi_s} \mathrm{d}\chi\, w(\chi, \chi_s)},\ee where $\tau$
is the magnification factor. Note that
$\eta$ is the average matter density relative to the global mean.
The variance of $\eta$ is given by (Valageas 2000a,2000b; Wang et
al. 2002) \be \label{eq:xieta} \xi_{\eta} = \int_0^{\chi_s}
\mathrm{d}\chi\, \left(\frac{w}{F_s}\right)^2\,I_{\mu}(\chi). \ee
Here \ba
F_s&=& \int _0^{\chi_s} \mathrm{d}\chi\, w(\chi, \chi_s),\nonumber\\
I_{\mu}(z)&=& \pi \int_0^{\infty} \frac{\mathrm{d}k}{k}\,\,
\frac{\Delta^2(k,z)}{k}\, W^2({\cal D}k\theta_0), \nonumber \ea
where $\Delta^2(k,z)= 4\pi k^3 P(k,z)$, $k$ is the wavenumber, and
$P(k,z)$ is the matter power spectrum. We used the non-linear power
spectrum from Peacock \& Dodds (1996), which is based on N-body
simulations. The matter power spectrum is shown in
Fig.\ref{mattersepectrum}. $W({\cal D}k\theta_0)=2J_1({\cal
D}k\theta_0)/ ({\cal D}k\theta_0)$ is the window function for
smoothing angle $\theta_0$. Because GRBs are point sources, we
adopted a sufficiently small smoothing angle $\theta_0=0.1''$ (Oguri
\& Takahashi 2006). Here $J_1$ is the Bessel function of the first
kind.

\begin{figure}
\includegraphics[width=9cm]{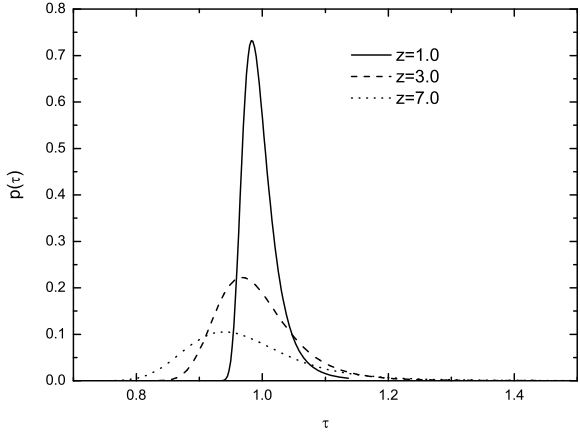}
\caption{Magnification probability distribution functions of
gravitational lensing at redshifts $z=1$, $z=3$ and $z=7$.
 \label{magnification}}
\end{figure}

Using $\tau =1+2|\kappa_{min}| (\eta-1)$ we can obtain \be p(\tau)=
\frac{g(\eta|\xi_{\eta})}{2|\kappa_{min}|}. \label{eq:mu,P(mu)} \ee
For an arbitrary cosmological model, one can compute $\xi_{\eta}$
from Eq.(14), and then the UPDF and $p(\tau)$ can be computed. In
Fig.\ref{magnification} we present the magnification probability
distribution functions $p(\tau)$ at redshifts $z=1$, $z=3$ and $z=7$
with $\Omega_M=0.27$ and $\Omega_{\Lambda}=0.73$. The results agree
with the $p(\tau)$ derived in Wang et al. (2002), Oguri \& Takahashi
(2006) and Holz \& Linder (2005). The probability distribution
functions at high redshifts have a higher variance and a lower
height of the maximum. The peaks reduce to a smaller magnification
factor $\tau$. From the probability distribution functions, we can
see that the gravitational lensing is crucial for high-redshift
objects.

\section{Constraints on cosmological parameters and dark energy including magnification bias}

\begin{figure}
\begin{center}
\includegraphics[width=9cm]{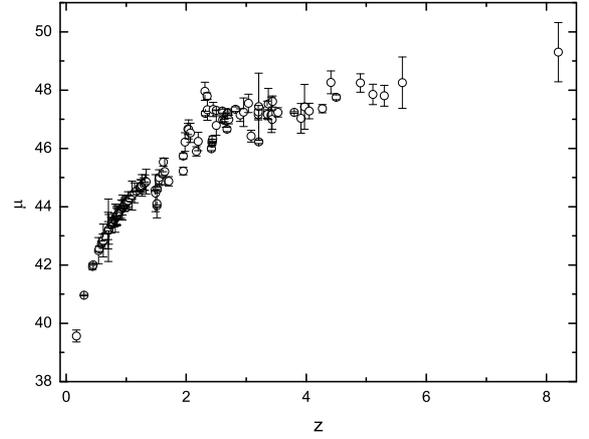}
\caption{Dispersions of GRBs distance modulus induced by
gravitational lensing.
 \label{lensing}}
\end{center}
\end{figure}
The random magnification of distant sources by the gravitational
lensing induces some bias in the observed sample. We calculated the
magnification bias as follows. Once the distribution $p(\tau)$ of
magnification factor $\tau$ is computed, the magnification bias is
drawn from the distribution $p(\tau)$. The dispersion of distance
modulus by the gravitational lensing is $\Delta \mu=-2.5$Log$\tau$.
In Fig.\ref{lensing} we show the dispersions of the GRB distance
modulus induced by the gravitational lensing. We can derive the
likelihood function from Bayes' rule (e.g., Gregory 2005). For each
point in a cosmological parameter space, the likelihood function is
determined by convolving the distribution of $\tau$ with the
intrinsic dispersion distribution (Dodelson \& Vallinotto 2006). We
used $h=0.742\pm0.036$ in our calculation (Riess et al. 2009).

\subsection{The $\Lambda$CDM cosmology}
\begin{figure}
\begin{center}
\includegraphics[width=9cm]{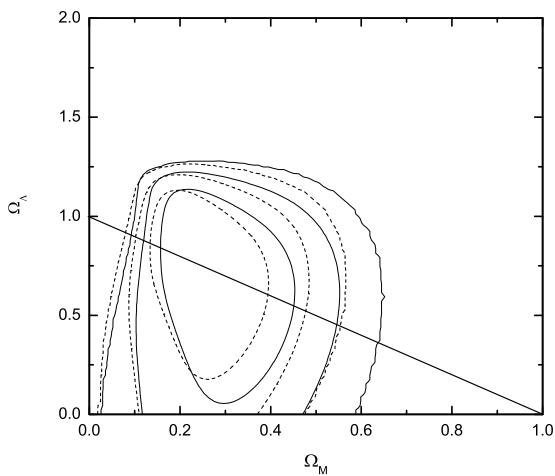}
\caption{Confidence contours of likelihood from $1\sigma$ to
$3\sigma$ in the $\Lambda$CDM model. The solid-line and dotted-line contours from
116 GRBs without and with the magnification bias.
 \label{CDM}}
\end{center}
\end{figure}

The luminosity distance in a Friedmann-Robertson-Walker (FRW)
cosmology with mass density $\Omega_M$ and vacuum energy density
(i.e., the cosmological constant) $\Omega_\Lambda$ is (Carroll,
Press \& Turner 1992)
\begin{eqnarray}
d_L & = & c(1+z)H_0^{-1}|\Omega_k|^{-1/2}{\rm sinn}\{|\Omega_k|^{1/2}\nonumber
\\ & & \times
\int_0^zdz[(1+z)^2(1+\Omega_Mz)-z(2+z)\Omega_\Lambda]^{-1/2}\}.
\end{eqnarray}
We used the GRB sample to constrain the cosmological parameters.
Fig.\ref{CDM} shows the $1\sigma$ to $3\sigma$ contour plotting in
the $\Omega_{M}-\Omega_{\Lambda}$ plane. The black line contours
from 116 GRBs show $\Omega_{M}=0.30_{-0.10}^{+0.09}$ and
$\Omega_{\Lambda}=0.84_{-0.78}^{+0.30}$ ($1\sigma$). The dashed
contours from 116 GRBs including gravitational lensing magnification
bias show $\Omega_M=0.26_{-0.09}^{+0.10}$ and
$\Omega_{\Lambda}=0.87_{-0.72}^{+0.26}$ ($1\sigma$). From the two
contours we can see that the gravitational lensing biases the
constraints on cosmological parameters. Because the solid line in
Fig.\ref{CDM} represents a flat universe, our result agrees with a
flat universe. The contours of GRBs at higher redshifts are almost
vertical to the $\Omega_M$ axis because the cosmology is
matter-dominated at high redshifts.

\subsection{The $w(z)=w_{0}$ model}
\begin{figure}
\begin{center}
\includegraphics[width=9cm]{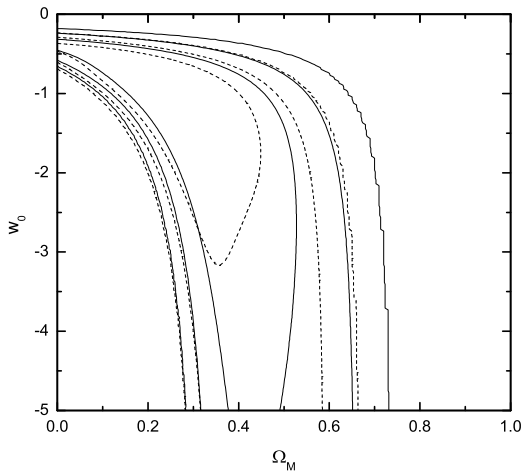}
\caption{Same as Fig.\ref{CDM} but in $w(z)=w_0$ model.
 \label{wcdm}}
\end{center}
\end{figure}

We consider an equation of state for dark energy
\begin{equation}
w(z)=w_{0}.
\end{equation}
In this dark energy model, the luminosity distance for a flat
universe is (Riess et al. 2004)
\begin{equation}
d_{L}=cH_{0}^{-1}(1+z)\int_{0}^{z}dz[(1+z)^{3}\Omega_{M}+(1-\Omega_{M})(1+z)^{3(1+w_{0})}]^{-1/2}.
\end{equation}
Fig.\ref{wcdm} shows the constraints on $w_{0}$ versus $\Omega_{M}$
in this dark energy model. The black solid line contours give
constraints from 116 GRBs and we have
$\Omega_{M}=0.29_{-0.28}^{+0.23}$($1\sigma)$ and $w_{0}=-1.1$. The
dashed contours give constraints from 116 GRBs including a
gravitational lensing magnification bias:
$\Omega_{M}=0.26_{-0.25}^{+0.18}$($1\sigma)$ and
$w_{0}=-1.05_{-2.03}^{+0.80}$.

\section{Model-independent constraints on the dark energy equation of state}
\begin{figure}
\begin{center}
\includegraphics[width=9cm]{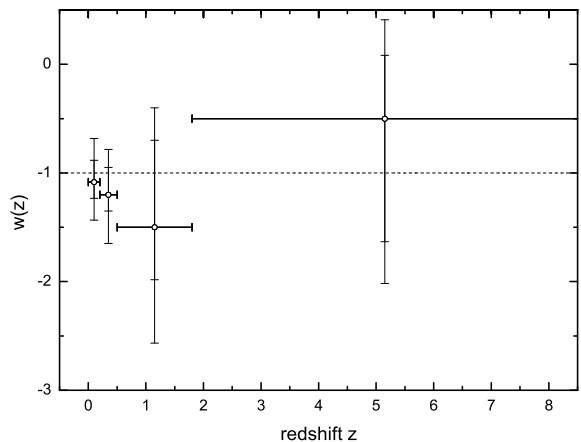}
\includegraphics[width=9cm]{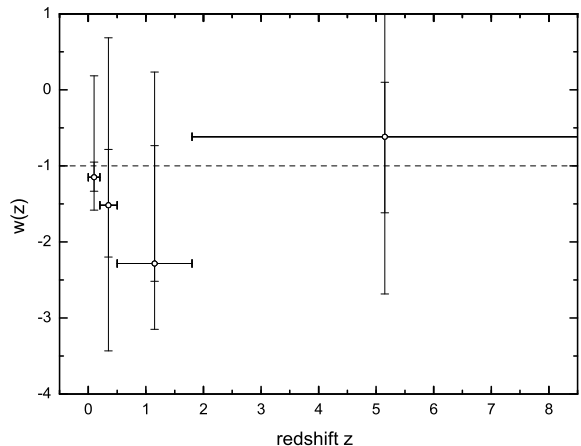}
\caption{Estimates of the uncorrelated dark energy EOS parameters
    $w(z)$.
    Top: uncorrelated dark energy parameters versus redshift, in which
    the vertical errorbars correspond to $1 \sigma$ and $2 \sigma$
    confidence levels of $w(z)$ and the horizontal
    errorbars span the corresponding redshift bins.
    Bottom: Same as top panel but including the weak lensing effect.}
    \label{wz}
\end{center}
\end{figure}

We first briefly describe a model-independent method to constrain
the equation of state (EOS) (for more details, see Qi, Wang \& Lu
2008a). We adopt the redshift binned parametrization for the dark
energy EOS as proposed in Huterer \& Cooray (2005), in which the
redshifts are divided into several bins and the dark energy EOS is
taken to be constant in each redshift bin but can vary from bin to
bin. For this parametrization, $f(z)$ takes the form (Sullivan et
al. 2007)
\begin{equation}
  \label{eq:fzbinned}
  f(z_{n-1}<z \le z_n)=
  (1+z)^{3(1+w_n)}\prod_{i=0}^{n-1}(1+z_i)^{3(w_i-w_{i+1})},
\end{equation}
where $w_i$ is the EOS parameter in the $i^{\mathrm{th}}$ redshift
bin defined by an upper boundary at $z_i$, and the zeroth bin is
defined as $z_0=0$. This parametrization scheme assumes less about
the nature of the dark energy, especially at high redshift, compared
with other simple parameterizations, because independent parameters
are introduced in every redshift range and it could, in principle,
approach any functional form with increasing the number of redshift
bins (of course, we would need enough observational data to
constrain all parameters well). For a given set of observational
data, the parameters $w_i$ are usually correlated with each other,
i.e. the covariance matrix
\begin{equation}
  \textbf{C}
  =\langle \textbf{w} \textbf{w}^{\mathrm{T}} \rangle
  - \langle \textbf{w} \rangle
  \langle \textbf{w}^{\mathrm{T}} \rangle
  ,
\end{equation}
is not diagonal. In the above equation, $\textbf{w}$ is a vector
with components $\widetilde{w_i}$ and the average is calculated by
letting $\textbf{w}$ run over the Markov chain. A new set of dark
energy EOS parameters $\widetilde{w_i}$ defined by
\begin{equation}
  \label{eq:transformation}
  \widetilde{\textbf{w}}=\textbf{T} \textbf{w}
\end{equation}
is introduced to diagonalize the covariance matrix. The
transformation of $\textbf{T}$ advocated by Huterer \& Cooray (2005)
has the advantage that the weights (rows of $\textbf{T}$) are
positive almost everywhere and localized fairly well in redshift,
which facilitates an interpretation of the uncorrelated EOS
parameters $\widetilde{w_i}$. The evolution of the dark energy with
respect to the redshift can be estimated from these decorrelated EOS
parameters. The transformation of $\textbf{T}$ is determined as
follows. First, we define the Fisher matrix
\begin{equation}
  \label{eq:fisher_matrix}
  \textbf{F}\equiv\textbf{C}^{-1}
  =\textbf{O}^{\mathrm{T}}\Lambda \textbf{O}
  ,
\end{equation}
where $\textbf{O}$ is orthogonal matrix and $\Lambda$ is diagonal.
Then the transformation matrix $\textbf{T}$ is given by
\begin{equation}
  \label{eq:transf_matrix1}
  \textbf{T}=\textbf{O}^{\mathrm{T}}
  \Lambda^{\frac{1}{2}}\textbf{O}
  ,
\end{equation}
except that the rows of the matrix $\textbf{T}$ are normalized such
that
\begin{equation}
  \label{eq:transf_matrix2}
  \sum_j T_{ij}=1
  .
\end{equation}

In addition to the Union2 SNe Ia sample and 116 GRBs, we also used
the distance ratio $D_V(0.35)/D_V(0.2)=1.736\pm0.065$ from SDSS7
data (Percival et al. 2010) and the shift parameter
$R=1.725\pm0.018$ from the WMAP seven-year data (Komatsu et al.
2010). The four redshift bins are $0-0.2$, $0.2-0.5$, $0.5-1.8$ and
$1.8-8.5$. We marginalize over Hubble parameter $H_0$, assuming a
broad uniform prior over the range $50<H_0<85$
km~s$^{-1}$~Mpc$^{-1}$. We also marginalize over $\Omega_M$ assuming
the quoted prior from Komatsu et al. (2010). Fig.~\ref{wz} shows the
constraints on EOS $w(z)$. This is the first time that the EOS is
constrained beyond the redshift $1.7$. From this figure we can
conclude that even though the EOS deviates from $\Lambda$ at
$1\sigma$ confidence level, it agrees with $w=-1$ at a $2\sigma$
confidence level.

\section{Discussion and conclusions}\label{Conclusions}
We have presented the gravitational lensing effects on constraints
of cosmological parameters and dark energy from GRBs. We mainly
focussed on the non-Gaussian nature of magnification probability
distribution functions and the magnification bias of gravitational
lensing. We first used an SNe Ia sample to calibrate the luminosity
relations of GRBs. Because the luminosity distances of SNe Ia are
completely cosmological-model-independent, the GRB luminosity
relations can be calibrated in a cosmology-model-independent way.
Then we calculated the PDFs of gravitational lensing. The
probability distribution functions at high redshifts have higher
variance and a lower height of the maximum. The peaks reduce to a
smaller magnification factor $\tau$. From the probability
distribution functions we can see that the gravitational lensing is
more important for high-redshift objects. Finally we presented
constraints on cosmological parameters and dark energy. We found
that the gravitational lensing had non-negligible effects on the
determination of cosmological parameters and dark energy. The
gravitational lensing shifts the best-fit constraints of
cosmological parameters and dark energy. Because high-redshift GRBs
are more likely to be reduced, the most probable value of the
observed matter density $\Omega_M$ is slightly lower than its actual
value. The gravitational lensing also biases a more negative value
of the dark energy equation of state. We also constrained the dark
energy equation of state out to redshift $z\sim 8$ in a
model-independent way using GRBs for the first time, and found that
the equation of state deviates from $\Lambda$CDM at the $1\sigma$
confidence level, but agrees with $w=-1$ at a $2\sigma$ confidence
level.

As shown in Samushia \& Ratra (2010), the cosmological constraints
from the two methods of Schaefer (2007) and Wang (2008) may be
different when using 69 GRBs. Therefore we emphasize that we need
detailed studies of new correlations with a much greater number of
GRBs and an examination of systematic errors to be able to regard
GRBs as more accurate standardizable candles. Now ongoing missions
like \emph{Swift}, \emph{Fermi} and \emph{Suzaku}, and the
collaboration of many observers on ground will promise the
progression of GRB cosmology.

\begin{acknowledgements}
We are grateful to Prof. Pengjie Zhang for fruitful discussion. This
work is supported by the National Natural Science Foundation of
China (grants 11103007, 10873009 and 11033002) and the National
Basic Research Program of China (973 program) No. 2007CB815404. FYW
is also supported by Jiangsu Planned Projects for Postdoctoral
Research Funds 1002006B and China Postdoctoral Science Foundation
funded projects 20100481117 and 201104521.
\end{acknowledgements}

\appendix

\section{Calculation of the lensing power spectrum}

The linear matter power spectrum $\Delta^2_L(k,z)$ is parameterized
as
\begin{equation}
\Delta^2_L(k,z)=Ak^{n_s+3}T^2(k)D^2(z), \label{linear}
\end{equation}
where $D(z)=g(z)/(1+z)g(0)$ is the linear growth factor, $T(k)$ is
the transfer function, $A$ is the normalization factor and $n_s$ is
the primordial fluctuation spectrum. We use the Harrison-Zel'dovich
spectrum $n_s=1$ throughout. For the $\Lambda$CDM model ($w=-1$),
the relative growth factor $g(z)$ is well approximated by (Carroll
et al. 1992)
\begin{equation}
g_{\Lambda}(z)=\frac{(5/2)\Omega_{M}(z)}{\Omega_{M}^{7/4}(z)-\Omega_{\Lambda}
(z)+(1+\Omega_{M}(z)/2)(1+\Omega_{\Lambda}(z)/70)},
\end{equation}
with
\begin{equation}
\Omega_{M}(z)=\frac{\Omega_{M}(1+z)^3}{\Omega_{M}(1+z)^3+\Omega_{\Lambda}},\
\Omega_{\Lambda}(z)=\frac{\Omega_{\Lambda}}{\Omega_{M}(1+z)^3+\Omega_{\Lambda}}.
\end{equation}
For the transfer function, we adopt the fitting result of Bardeen et
al. (1986) for an adiabatic $\Lambda $CDM model
\begin{equation}
T_{\Lambda}(q) = \frac{\ln(1+2.34q)}{2.34q}[1+3.89q+(16.1q)^2+
(5.46q)^3+(6.71q)^4]^{-1/4},
\end{equation}
where $q=k/h\Gamma$, and $h=H_0/(100 \,{\rm km}\,{\rm s}^{-1}\, {\rm
Mpc}^{-1})$,and $\Gamma=\Omega_{M}h\exp[-\Omega_{b}(1+\sqrt{2h}/
\Omega_{M})]$ is the shape parameter with baryon density
$\Omega_{b}$.

For the non-linear power spectrum we adopt the formula given by
Peacock \& Dodds (1996),
\begin{eqnarray}
\Delta_{NL}^2(k_{NL})&=&f_{NL}[\Delta_L^2(k_L)],\nonumber\\
k_L&=&[1+\Delta_{NL}^2(k_{NL})]^{-1/3}k_{NL},\nonumber\\
f_{NL}(x)&=&x\left[\frac{1+B\beta x+(Ax)^{\alpha\beta}}
{1+[(Ax)^{\alpha}g^3(z)/(Vx^{1/2})]^{\beta}}\right]^{1/\beta}.
\end{eqnarray}
The parameters in the non-linear function $f_{NL}$ are
\begin{eqnarray}
A&=&0.428(1+n_s/3)^{-0.947},\nonumber\\
B&=&0.226(1+n_s/3)^{-1.778},\nonumber\\
\alpha&=&3.310(1+n_s/3)^{-0.244},\nonumber\\
\beta&=&0.862(1+n_s/3)^{-0.287},\nonumber\\
V&=&11.55(1+n_s/3)^{-0.423},\nonumber
\end{eqnarray}
which are fitted to the numerical simulation results.

\end{document}